%% file: e6_1_staggs_1023.tex
\documentclass{revtex4}

\usepackage{graphicx}
\usepackage{e6_1_staggs_1023_aas_macros}
\input epsf

\def\apjl{Astrophys. J. Lett.}

\def\be{\begin{equation}}
\def\ee{\end{equation}}
\def\bea{\begin{eqnarray}}
\def\eea{\end{eqnarray}}

\def\cmm2{{\,\rm cm^{-2}}}
\def\cm2{{\,{\rm cm}^2}}
\def\cmm3{{\,{\rm cm}^{-3}}}
\def\gcmm3{{\,{\rm g\,cm^{-3}}}}

\def\fun#1#2{\lower3.6pt\vbox{\baselineskip0pt\lineskip.9pt
  \ialign{$\mathsurround=0pt#1\hfil##\hfil$\crcr#2\crcr\sim\crcr}}}

\def\p3m{P$^3$M}

\def\ga{\mathrel{\mathpalette\fun >}}
\def\fun#1#2{\lower3.6pt\vbox{\baselineskip0pt\lineskip.9pt
  \ialign{$\mathsurround=0pt#1\hfil##\hfil$\crcr#2\crcr\sim\crcr}}}

\begin{document}
\bibliographystyle{e6_1_staggs_1023_prsty}


\title{Experimental CMB Status and Prospects:  A Report from Snowmass 2001}


\author{Suzanne Staggs}
\affiliation{Princeton University}
\email{staggs@princeton.edu}
\author{Sarah Church}
\affiliation{Stanford University}
\email{schurch@stanford.edu}


\date{October 15th, 2001}

\begin{abstract}
In recent years the promise of experimental study of
the cosmic microwave background (CMB) has been demonstrated.
Herein a brief summary of the field is followed by an indication
of future directions.  The prospects for further revelations from the
details of the CMB are excellent.  The future work is well-suited to the
interests of high energy physics experimentalists:  the projects
seek to understand the fundamental nature of the universe,
require subtle experimental techniques to keep systematics in
control, produce large, rich data sets, and must be implemented by
multi-institutional collaborations.
\end{abstract}

\maketitle


\section{Introduction}
\label{intro}
\input{e6_1_staggs_1023_intro}
\section{Current Status of Temperature Anisotropy Measurements}
\label{temp_summary}
\input{e6_1_staggs_1023_temp}
\section{CMB Polarization Experiments and Techniques}
\input{e6_1_staggs_1023_pol}
\label{polarization}
\section{Fine-Scale CMB and Sunyaev-Zel'dovich Experiments}
\input{e6_1_staggs_1023_fscmb}
\label{sz}
\section{Conclusions}
\label{conclusions}
\input{e6_1_staggs_1023_conc}


%
%

%
%

\begin{acknowledgments}
\input{e6_1_staggs_1023_ack}
\end{acknowledgments}

\bibliography{e6_1_staggs_1023}

\end{document}

%% file: e6_1_staggs_1023_intro.tex
The vast cache of information encoded in the CMB's intensity 
distribution has been tapped in recent years.  In the last three 
years, a spate of independent experiments has evinced 
the primordial fluctuations present when the radiation decoupled from 
the matter at a redshift of $z\approx 1100.$  Continuing measurements 
of the CMB temperature anisotropy should soon be joined by measurements 
of its polarization anisotropy.  The polarization presents the enticing 
possibility of educing information about the inflation field itself.
A third exciting area of CMB research is the measurement of its 
fine-scale anisotropy, at $\ell\ga 2500$, where secondary effects 
dominate, so that the CMB serves as a backlight to illuminate dark 
parts of the universe. One of the main features of the fine-scale 
anisotropy is the Sunyaev-Zel'dovich effect in clusters, causing 
clusters to appear as either hot or cold spots, depending on the 
frequency of observation.

%% file: e6_1_staggs_1023_temp.tex
\begin{figure}[htbp]
  \includegraphics[width=0.65\columnwidth]{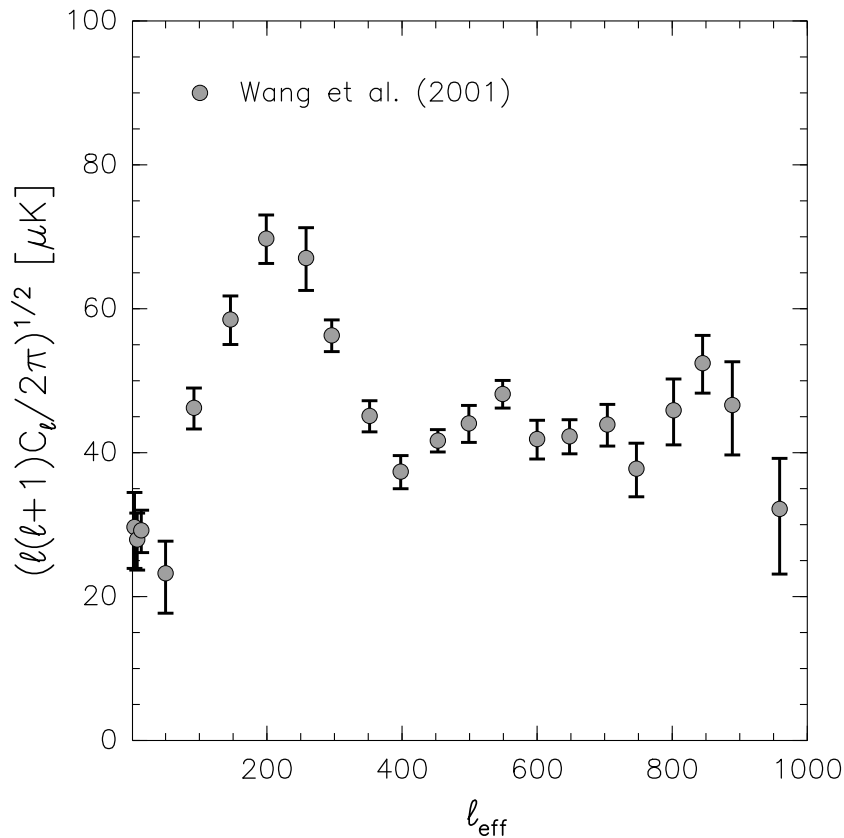}
  \caption[]{The status of measurements of the temperature anisotropy
  angular power spectrum of the CMB as of May, 2001, based on the
  compilation of \protect{\cite{wtz01}}.  The authors binned 105 measurements of
  the power spectrum, including the effects of beam and calibration
  uncertainties.  The individual measurements came from a variety of
  experiments, detailed by Wang, Tegmark \& Zaldarriaga.  Figure
  courtesy of M. Nolta.}
  \label{fig:wang}
\end{figure}

Recent stunning results from the CMB have been measurements of its
temperature anisotropy at angular scales between about $0.\!\!^\circ 2$
and $2^\circ$.  Data as of May, 2001, are plotted in the angular power
spectrum of Figure~\ref{fig:wang}, adapted from the compilation of Wang,
Tegmark \& Zaldarriaga (2001)\cite{wtz01}.  The power spectrum clearly
shows a peak near $\ell=200$, and is consistent with the multiple peaks
expected from oscillations of the primordial plasma at the time the
radiation decoupled from matter. Much of the excitement generated by these
data springs from the hope of fitting the measured power spectrum to
determine fundamental cosmological parameters.  An indication of the
consistency of some of the recent larger data sets is provided in
Table~\ref{table:knox}, where parameters are estimated based on individual
data sets.

\begin{table}[htbp]
\caption[]{Published results from fitting parameters to several recent large data
sets, under the assumption of no gravitational waves.  The table is
based on one compiled by L. Knox. }
\label{table:knox}
\begin{tabular}{|l|l|l|l|l|l|l|} \hline
Experiment & $\Omega_{b}h^{2}$ & $\Omega_{cdm}h^{2}$ & $\Omega_{tot}$
& $n_{s}$ & Prior & Reference \\
\hline\hline
DASI & $.022^{+.004}_{-.003}$ & $.14\pm.04$ & $1.04\pm.06$ &
$1.01^{+.08}_{-.06}$ & $h>0.45$; $\tau=0$ & \cite{pryke01}\\
BOOM & $.022^{+.004}_{-.003}$ & $.13\pm.05$ & $1.02\pm.06$ &
$0.96^{+.10}_{-.09}$ & $0.4<h<0.9$ & \cite{net01}\\
Maxima & $.030^{+.009}_{-.005}$ & $.20^{+.10}_{-.05}$ & $1.08^{+.16}_{-.04}$
& $1.00^{+.12}_{-.15}$ & $0.4<h<0.9$ & \cite{balbi00,balbi01}\\
\hline
\end{tabular}
\end{table}

\subsection{Planned and Ongoing Temperature Anisotropy Experiments.}
\begin{figure}[htbp]
  \includegraphics[width=0.65\columnwidth]{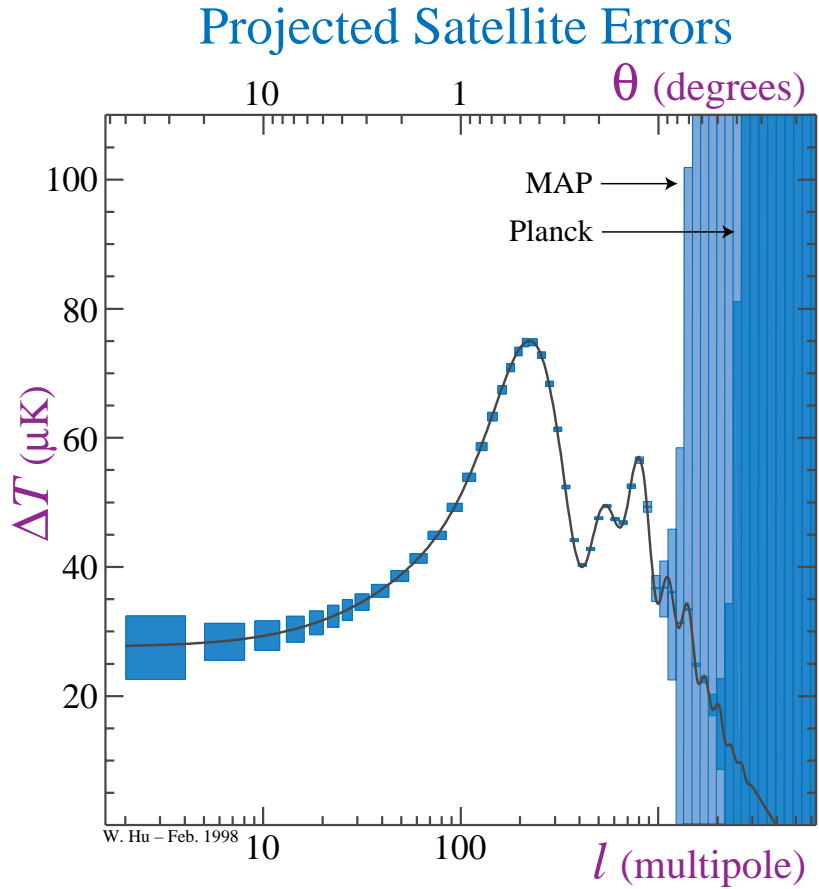}
  \caption[]{Projected errors on the temperature anisotropy angular
  power spectrum for the CMB.  Error boxes shown superimposed on
  a specific realization of the power spectrum represent estimates of
  the $1\sigma$ (68\%
  CL) errors for measurements made at a given $\ell$.  The width of
  each box indicates the width of the ``window function'' used for
  averaging the data in that $\ell$-bin. Figure
  courtesy of W. Hu.}
  \label{fig:map}
\end{figure}

On the day before the Snowmass 2001 meeting began, on 30 June 2001,
the Microwave Anisotropy Probe (MAP) was launched into space by NASA.  The
satellite has since reached its orbital position at L2, the saddle
point four Earth-moon distances beyond the moon.  Though by prior
arrangement no results will be released for another year, the science
team~\cite{pagepri} reports that all systems are functioning well. The eponymous MAP
will provide full-sky maps of the CMB fluctuations with high
signal-to-noise measurements of the temperature power spectrum out to
$\ell\approx 1000$, as indicated by the error boxes in
Figure~\ref{fig:map}.  The MAP data will reflect the advantages of a
space mission:  extreme control of systematic effects is possible, and
the entire $4\pi$ of the sky can be measured.

In 2007, a second satellite will be launched: the Planck Surveyor which is
a joint ESA/NASA mission. Planck will measure temperature fluctuations
with high signal-to-noise out to about $\ell\approx 2000$, and will also
measure polarization, as discussed below. See Figure~\ref{fig:map} for
estimates of Planck's sensitivity. It seems likely that after Planck, no
other measurements of the CMB temperature power spectrum at $\ell < 2000$
will be needed.

Several other experiments are ongoing:  DASI continues to collect
data from the South Pole (though retooled to measure all four Stokes
parameters now); Boomerang and Maxima each plan to fly again, to
measure polarization as well as remeasure temperature anisotropy.
CBI and ACBAR~\cite{holzpri} have collected additional data, and  MINT
(a 140~GHz SIS-mixer-based four-element interferometer~\cite{pagepri})
is presently deployed on the Chilean altiplano.

\subsection{Concluding Remarks about the Temperature Anisotropy}
The future looks bright for collecting definitive data on the
power spectrum of the CMB out to $\ell\approx 2000$.  Furthermore,
large-scale, high-resolution maps of the CMB crucial for looking
for nongaussian effects will soon be available.  Published data
already show good agreement among CMB experiments and consistency
when pitted against other measurements of cosmological parameters.
The days of making do with exiguous data are past.

%% file: e6_1_staggs_1023_pol.tex
The polarization of the CMB reveals more information about the
universe than the temperature anisotropy can alone.  However, the
polarization is predicted to be about twenty times smaller than the
temperature anisotropy.  To date, no polarization anisotropy has been
detected, though such detections should be imminent if rough theoretical
predictions for its magnitude are correct (based on the sensitivities
of ongoing and planned experiments).

The reader is referred to the report from the Snowmass working group P4
\cite{cjk01} for more details on the theoretical prospects for
interpreting polarization data (as well as references for further
reading). The polarization field is a tensor field which can be decomposed
into a ``curl'' part and a ``gradient'' part, referred to as the $B$ modes
and the $E$ modes, respectively. It turns out that the acoustic density
and velocity perturbations which give rise to the bulk of the temperature
anisotropy signal produce only $E$ modes, while gravitational waves from
the inflation field itself can generate both $E$ and $B$ modes.  Thus,
detection of $B$ modes could constrain inflation in a fundamentally new
way.

\begin{figure}[htbp]
   \includegraphics[width=4in]{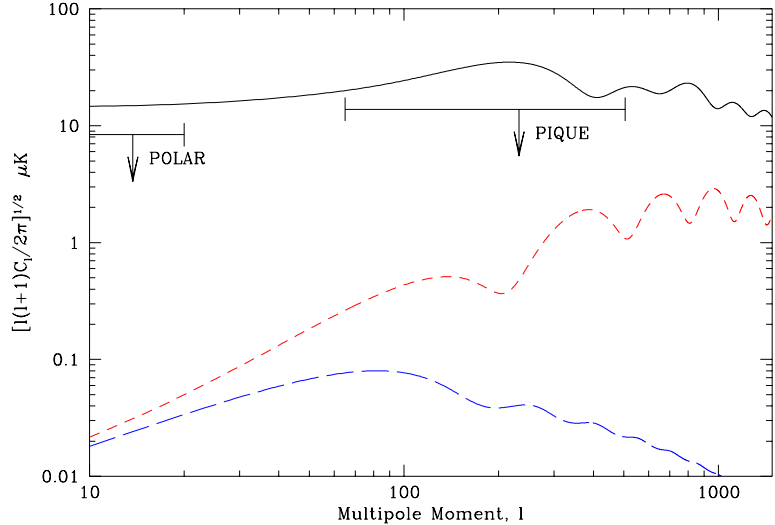}
   \caption[]{Predictions for the power spectrum of the polarization
    of the CMB made using the CMBFAST code~\protect{\cite{sel96}}.
    The assumed parameters in the model are $\Omega_b=0.05$, $\Omega_{dm}$=0.25, $\Omega_\Lambda=0.7$ $n_S=0.94$,
$n_T=0.94$, $Q_T/Q_S=0.28$. Also shown are the best upper limits at
$l<1000$: POLAR\cite{POLAR01} and PIQUE\cite{PIQUE01}.}
   \label{pol-fig1}
\end{figure}

Many of the techniques that have been used to detect temperature
anisotropies in the CMB can be directly
applied to measurements of CMB polarization.  Additionally,
polarization techniques from radioastronomy may be adapted.
However, though the amplitude of
{\em E}-mode polarization fluctuations is expected to be ``only''
twenty times smaller than the temperature anisotropies, the {\em B}-modes
are likely to be ten times smaller still (see Fig.~\ref{pol-fig1}).
Consequently, not only must the ultimate polarization experiments be
of order 100 times
more sensitive than current anisotropy experiments, they must also
have 100 times better rejection of systematic signals.

\subsection{How to Measure Polarization}
Rather than $E$ and $B$, which are global properties of the radiation
field, experiments measure the Stokes parameters describing its linear
polarization:  $Q$ and $U$. (No circular polarization is predicted from
primordial effects.)  Of course, $Q$ and $U$ depend on the coordinate
system chosen by the experimenters. Most experiments will require some
modulation of the instrument coordinate system with respect to the sky
(``chopping'') to reduce systematic effects due to common-mode pick-up.
Fortunately, as in the temperature case, prescriptions have been developed
for using likelihood analysis on chopped $Q$ and $U$ data to constrain the
power spectra of $E$ and $B$.  (See \cite{mz98} and references therein,
for example).

\subsection{Detection Techniques}
Just as with temperature anisotropy measurements,
polarization measurement techniques fall in two types -- coherent
detection, in which the individual photons are amplified before detection,
and incoherent detection, in which the photons are directly detected
without amplification. Typically, measurements below 100~GHz are made
with coherent techniques (high electron mobility amplifiers, or HEMTs), while those
above are made with incoherent methods (bolometers).  At 100~GHz,
both techniques are feasible. At 150~GHz, SIS mixers provide another
method of coherent amplification.
The noise limitations of the two techniques are
fundamentally different (see for example, \cite{zmu99}):  bolometers
are intrinsically more sensitive (and even more sensitive when used in a space
environment with cooled optics).
However, coherent techniques for
rejecting large common-mode signals (ie, the intensity itself in this case) are
well-understood from interferometry work.   Also, HEMTs can
be operated at temperatures of 4-20K allowing for relatively simple
cryogenic systems based on mechanical cryocoolers. Bolometers need to
be cooled to $300$~mK or lower to realize high  sensitivities.
A final consideration is the problem of foregrounds:  at frequencies
below 100~GHz, polarized synchrotron radiation (with $T\propto
\nu^{-\alpha}$ where $\alpha \sim 2.7$) is the primary
contaminant, and above 100~GHz, polarized dust emission dominates.

\subsection{Experimental Configurations}
\label{pol-sec3}
Both coherent and incoherent detection techniques are playing an important
r\^ole in the search for CMB polarization.  There are three types of
experiments that are currently operating, or are under construction.
The first two types (correlation polarimeters and interferometers)
rely on phase-coherent correlations and thus are easiest to implement
with coherent devices.

\begin{figure}[htbp]
    \includegraphics[width=3.5in]{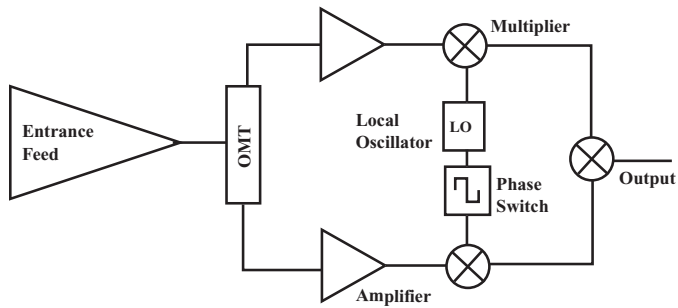}
    \caption[]{Schematic of a correlation polarimeter.  The phase
    switch in the local oscillator path multiplies the output by $\pm 1$
    at kHz rates, well above the $1/f$ noise of the amplifiers.  The
    output is proportional to the Stokes parameter $U$ if the OMT
    axes define $x$ and $y$.}
    \label{pol-fig3}
\end{figure}

{\bf Correlation Receivers} The best upper limits so far on CMB
polarization have been obtained by the PIQUE~\cite{PIQUE01} (at 90~GHz)
and
POLAR~\cite{POLAR01} (at 30~GHz) experiments (see Fig.~\ref{pol-fig1}). Both
experiments use heterodyne correlation polarimeters, which
employ phase-coherent techniques similar to those in interferometers.
The key to the correlation polarimeter is that its output is
directly proportional to one linear Stokes parameter:  it does not
require differencing of two large signals to a few parts in $10^{7}$.
One such polarimeter is
shown schematically in Fig.~\ref{pol-fig3}. The
signal from the sky is split into two orthogonal linear polarization
states by an orthomode transducer (OMT). Each polarization state is then
amplified by a HEMT amplifier. The radio frequency
signals are down-converted to a lower frequency by mixing with a
local oscillator.  The two
orthogonal polarization states are then multiplied together at the output
of the receiver. Note that if the input axes are labeled $x$ and $y$,
then this polarimeter measures $U$, the linear polarization measured
with respect to axes rotated by $45^\circ$ to the $x$-$y$ system.

The purpose of the phase switch, which periodically switches the phase of
one branch of the LO between 0 and $\pi$, is to periodically reverse the
sign of the output.  This switching can take place at several kHz
to modulate the output well above the $1/f$ noise from the
amplifier.  The result is extremely stable receivers:  the $1/f$ knee
of the PIQUE polarimeter was undetectable with measurements out to
time scales of days.


HEMT technology is well-established and correlation receivers
also have long track records in the radioastronomy community.
The PIQUE and POLAR experiments are the first to demonstrate that
systematic effects can be controlled to the few $\mu$K level for
polarization measurements.  However,
the sensitivity of a coherent receiver is limited by the noise temperature
of the amplifiers which cannot be lower than the quantum limit,
$T=h\nu/k$.  Consequently, at 100 GHz, for instance, a coherent receiver
must add at least 5K of noise to the system. In practice, the best
coherent receivers have noise temperatures about twice the quantum limit.
The sensitivity of a coherent receiver may be quoted in terms of a noise
equivalent temperature (NET; usually expressed in
$\mu\mbox{K}\sqrt{\mbox{s}}$) which is given by:
\be {\rm NET} = f \frac{\sum_i  T_i}{\eta\sqrt{\Delta\nu}},
\label{pol-eq6}
\ee
where the sum includes
Rayleigh-Jeans equivalent temperatures for the CMB, the atmosphere,
and the receiver (and any other incident noise).
The factor $f$ depends on details of the receiver; for a correlation
receiver, $f=\sqrt{2}$.  Here,
$\Delta\nu$ is the bandwidth of the receiver and $\eta$ is the system
efficiency (typically $>0.95$ for HEMT-based systems).
For ground-based experiments, HEMT's are typically half as
sensitive as bolometers.  (Balloon-borne bolometers can be more than
five times as sensitive as ground-based HEMTs, and the improvements in
sensitivity for deployments in space are even more impressive.)
A challenge for future
experiments based on HEMTs will be to build the large focal
plane arrays that will be needed to achieve the sensitivity required to
detect $B$ mode polarization.  One avenue might be through monolithic
microwave integrated circuit (MMIC) antenna-coupled HEMT correlation
receivers.

\begin{figure}
    \includegraphics[width=0.6\columnwidth]{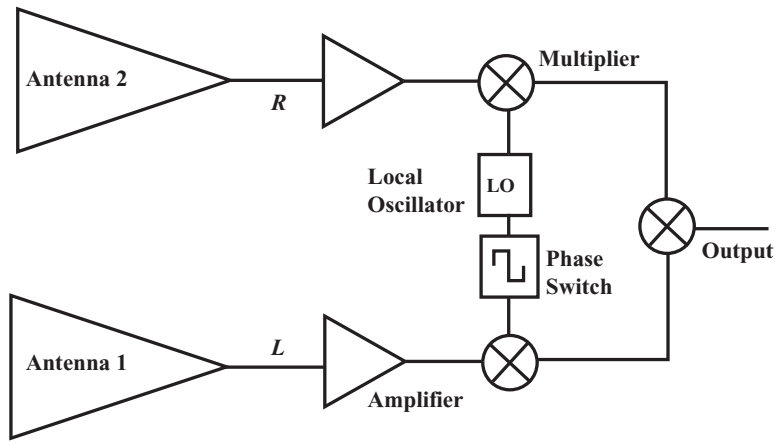}
    \caption[]{Schematic of the correlation of one pair of antennas in
    an interferometer, meant to emphasize its similarity to the
    correlation polarimeter of Figure~\ref{pol-fig3}}
    \label{pol-fig4}
\end{figure}

{\bf Interferometric Techniques}  Some of the recent new data on the CMB
temperature anisotropy has come from two interferometers operating in the
30-40~GHz range:  DASI \cite{lei01} and CBI \cite{padin01} . Both have now
been reconfigured to make polarization measurements. As shown in
Fig.~\ref{pol-fig4}, the instrumentation is similar to that in
Fig.~\ref{pol-fig3}, but now the two inputs come from two antennas, rather
than from the two arms of an OMT. Typically a single circularly polarized
component of the radiation is amplified from each antenna.  The outputs
from each pair of antennas are correlated (multiplied together).
Consequently the number of correlator channels required is proportional to
$n^{2}$ where $n$ is the number of antennas in the array. Inteferometers
provide some immunity to local sources of systematics, including
atmospheric emission and warm ground spillover. Like correlation
polarimeters, when configured to measure polarization, interferometers do
not have to difference two large signals to high accuracy; the correlation
outputs are directly proportional to (linear combinations of) linear
Stokes parameters. Interferometers also {\em directly measure} the power
spectrum of fluctuations because each baseline is sensitive only to a
narrow range of Fourier components of the sky brightness and thus to a
narrow range of multipoles.

The sensitivity of an interferometer can be calculated from
Equation~\ref{pol-eq6} with:

\[ f=\frac{\Omega_p}{\Omega_s}\frac{1}{\sqrt{n(n-1)}} \]

where $\Omega_p$ and $\Omega_s$ are the solid angles of
a single antenna's beam (the field of view) and
the synthesized beam, respectively.

Typically, to make polarization measurements, each antenna is configured
to measure one component of {\em circular} polarization -- either left
($L$) or right ($R$). Quarter-wave plates can be used to periodically
switch each antenna between $L$ and $R$ states.  This switching technique is
implemented in
DASIPOL~\cite{jcpri}.  All possible
correlations between a pair of antennas -- $LR$,
$LL$, $RR$ and $RL$ -- can be measured, from whence all
four Stokes parameters may be determined, including $V$, the
degree of circular polarization.  The ability to measure $I$, $Q$ and $U$
quasi-simultaneously confers an important advantage to the interferometric
technique. Interferometers can measure the temperature, polarization,
and $TE$ cross-correlation power spectra all with the same detector
set.  The primary disadvantage of an interferometer is that
scaling it up to have a large number of detectors is challenging. The
number of correlations grows as $n^2$ while the sensitivity grows as
$n$. The
state-of-the-art DASI and CBI instruments each have thirteen
elements; to get ten times more sensitivity by a brute-force
scaling up is impractical.

{\bf Direct Detection with Bolometers.} Bolometers have
been used with great success to measure the CMB anisotropy power spectrum as
part of the balloon-borne Boomerang~\cite{bern00} and
MAXIMA~\cite{han00}
experiments.  Both are being
reconfigured to observe polarization, and a
number of other experiments that will use bolometers to measure
polarization are also under construction (see Table~\ref{pol-t1}).

\begin{figure}
    \includegraphics[width=3in]{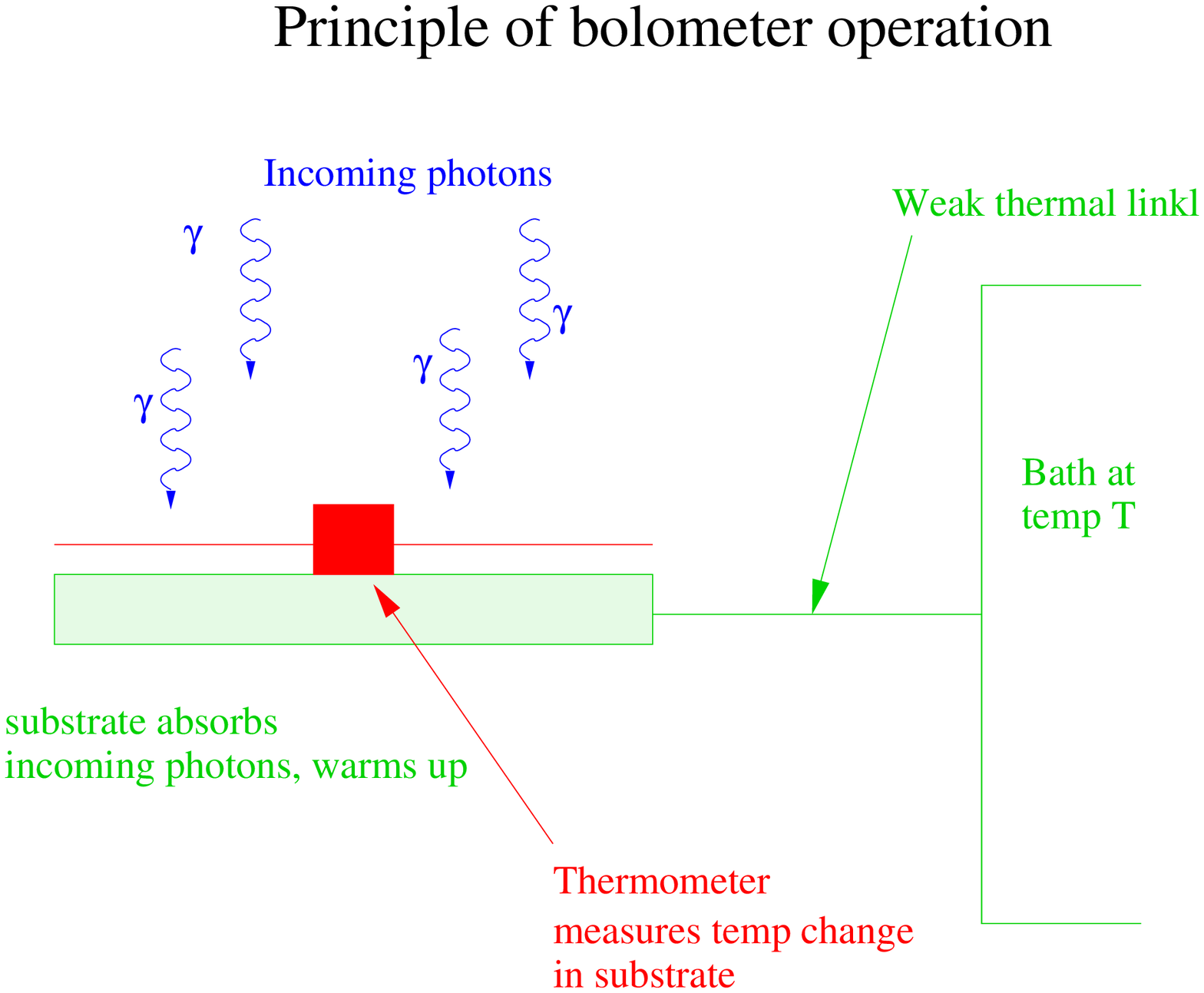}
    \includegraphics[width=3in]{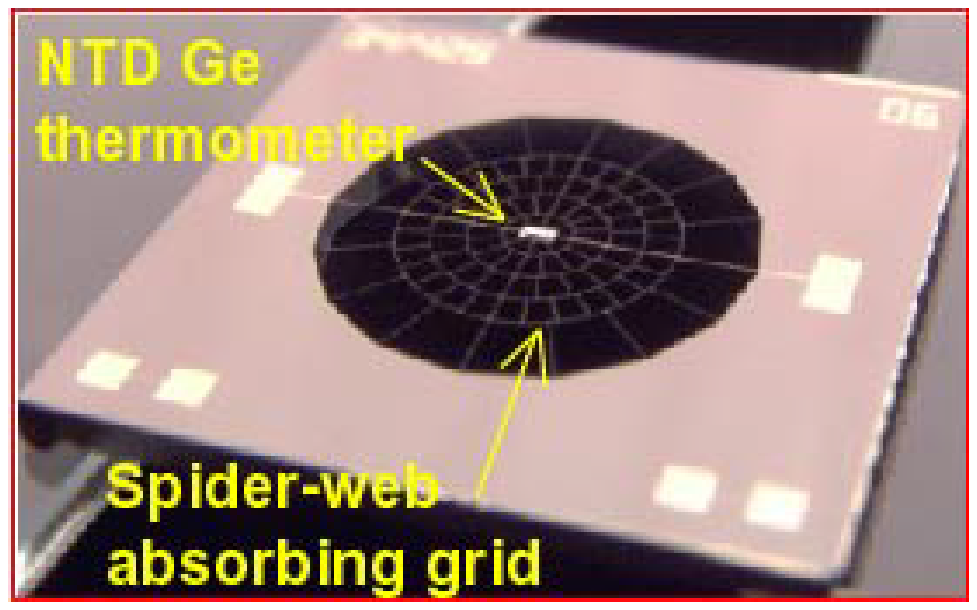}
    \caption[]{Left: Principle of bolometer operation.  Right: Picture of a
silicon-nitride ``spider-web'' bolometer [Courtesy James Bock, Jet
Propulsion Laboratory].}
    \label{pol-fig5}
\end{figure}

A bolometer comprises a thermistor
attached to a substrate capable of absorbing photons (see
Fig.~\ref{pol-fig5}). The photons heat the substrate, causing a
temperature rise that is proportional to the absorbed power. The bolometer
show in the lower panel of Fig.~\ref{pol-fig5} has a substrate made of
silicon-nitride etched into a ``spider-web'' pattern.  This process
removes most of the material, creating a sensitive, low heat-capacity
structure, yet retains high optical efficiency because the gaps in the
substrate are much less than a wavelength.  The thermistor shown in the
picture is neutron transmutation-doped germanium (NTD-Ge) which behaves as
a temperature dependent resistor.

The sensitivity of a bolometer receiver may be described by a noise
equivalent power (NEP; usually quoted in W/$\sqrt{\mbox{Hz}}$),
which depends on the details of the
thermistor, the substrate, the temperature and stability of
the bath, the thermal coupling to the bath, and the readout electronics.
In general, $\mbox{NEP}^2 = \sum_i\mbox{NEP}_i^2$, where the index
$i$ enumerates different sources of noise, including photon noise,
Johnson noise,  phonon noise, and other sources. (See the
seminal work by John Mather: \cite{mather82,mather84}.)
Bolometers may be designed, constructed and operated such
that the photon NEP dominates their noise;
in that case they are said to have background limited performance
(BLIP).  The BLIP NEP for a bolometer with several sources $P_i$ of
power incident on it is given by:
\[ {\rm NEP}_{\rm BLIP}=\sqrt{\sum_i[2P_i
(h\overline{\nu}+\epsilon_i\eta_i k_B T_i)]}, \]
where $\overline{\nu}$ represents the average frequency in the pass-band.
Power sources incident on the detector may include the atmosphere, warm
emission from the telescope and surroundings, and the CMB. The first term
in parentheses is the contribution from random photon arrivals (shot noise
due to Poisson statistics). The second term accounts for the effect of
photon (boson) correlation and depends on the source emissivity
$\epsilon_i$ and temperature $T_i$, and on the net efficiency through the
optics to the detector, $\eta$. The NEP can be converted to an NET for
comparison with a coherent receiver as follows:

\[ {\rm NET} = \frac{1}{\sqrt{2\eta}}
\frac{{\rm NEP}}{(\partial P_{\rm CMB}/\partial T_{\rm
CMB})}\,, \]

\noindent where $\eta$ is the detection efficiency for CMB photons and
$P_{\rm CMB}$ comes from the Planck radiation law.  For example, the
300~mK spiderweb bolometers used by Boomerang \cite{pia01}
had typical NEPs of $3\times 10^{-17}\mbox{ W}/\sqrt{\mbox{Hz}}$,
corresponding to a $\mbox{NET}_{\rm CMB}\approx
200\mu\mbox{K}\sqrt{\mbox{s}}$.  The best Maxima bolometers (100~mK)
had  $\mbox{NET}_{\rm CMB}\approx
100~\mu\mbox{K}\sqrt{\mbox{s}}$ \cite{lee99}.
For comparison, the best QMAP HEMT-based
receiver had a sensitivity of $400~\mu\mbox{K}\sqrt{\mbox{s}}$
\cite{mil01}.

Bolometers are not intrinsically
sensitive to the polarization state of the incoming radiation. Linear
components of the radiation field can be selected by (i) placing wire-grid
polarizers in front of the bolometer, (ii) using an OMT and then detecting
the two linear polarization modes with separate bolometers, or by (iii)
making the bolometer substrate itself polarization sensitive by a suitable
choice of geometry for the substrate. Future experiments will use all
three methods.

Bolometers respond equally well to {\em all} frequencies.  Consequently
the radiation must be carefully filtered before it reaches the bolometer.
Fig.~\ref{pol-fig6} shows how a bolometer is coupled to a telescope.
Metal-mesh resonant grid filters define a pass-band with high transmission
and low out of bands leaks~\cite{leec96}.  Typically the structure shown
in the figure has a transmission efficiency of 40\%~\cite{church96}.

\begin{figure}
   \caption[]{Schematic showing how a bolometer is coupled to the sky.}
   \label{pol-fig6}
\end{figure}

\begin{figure}
    \includegraphics[width=4in]{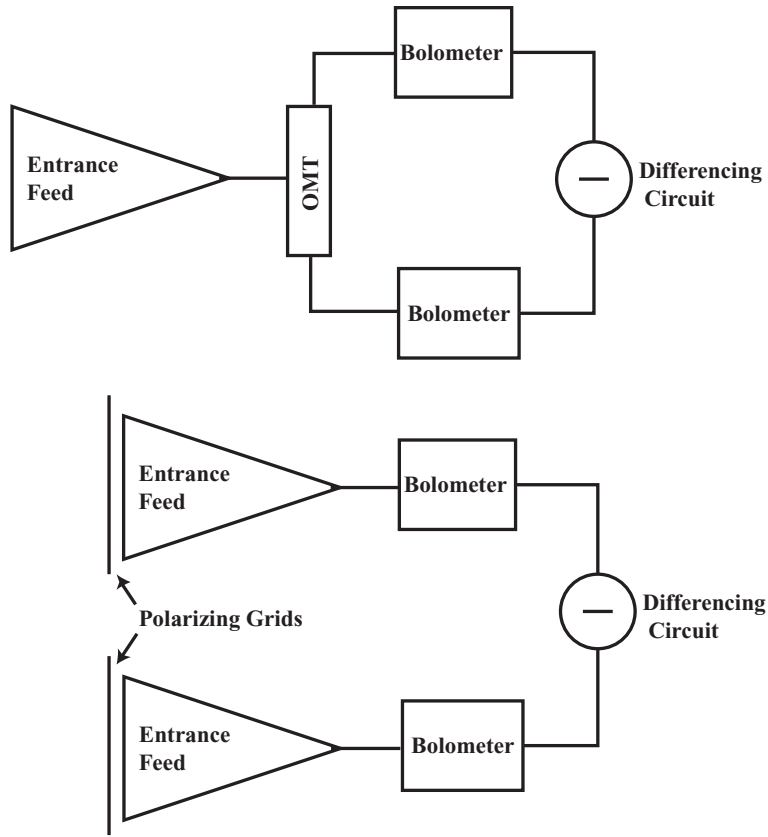}
    \caption[]{Two ways in which bolometers can be used to measure
polarization.}
    \label{pol-fig7}
\end{figure}

Figure~\ref{pol-fig7} shows schematically how this structure can be
adapted to measure polarization. In the top panel, an OMT is used to split
the signal into two orthogonal components which are
then detected by separate bolometers~\cite{phil01}. In the bottom panel,
two separate feeds are used to measure orthogonal components
by placing a wire polarizing grid in front of each feed.

To measure $Q$ and $U$, the outputs of two bolometers are
differenced. Pairs of bolometers
must be well-matched to allow rejection of the common-mode signal
which is $10^{6-7}$ times larger than the signals of interest.
(Alternatively, mechanical rotation of either a grid or a quarter-wave
plate allows linear Stokes parameters to be measured by a single
bolometer; these techniques have their own technical problems.)
The SuZIE measurements have successfully differenced matched pairs
of bolometers \cite{hol97}.  If the instrument coordinate
system is at an angle
$\theta$ to the sky coordinate system, the output from the receiver is
proportional to $[Q\cos 2\theta + U\sin 2 \theta]$.  Rotating the
receiver thus provides good systematic checks. Ideally, to minimize
systematics, the two bolometers should simultaneously view the same pixel
on the sky. Although this can be achieved by using OMTs, the resulting
architecture is bulky, limiting the number of feeds that can be packed
into a focal plane. A more compact method recently developed replaces the
dual-polarization bolometer in Figure~\ref{pol-fig6} with a
polarization sensitive bolometer (PSB): two
vertically stacked bolometers each sensitive to one of two orthogonal
linear components of the radiation.  Both approaches will soon be
tested -- the Polatron experiment uses an OMT~\cite{phil01}, and a new
flight of Boomerang will provide the first data from PSBs.

There are good prospects for developing large focal plane arrays of
bolometers.  BOLOCAM \cite{bolocam} already comprises 151 detectors.
The NTD Ge thermistors can be replaced with
photolithographed transition edge superconductor~\cite{lee96,Benford00}.
Recent submm instruments based on ``pop-up'' detectors (bolometers
whose leads are folded down out of the focal plane to allow
close-packing) have demonstrated new multiplexing techniques
\cite{Benford01,Benford00} which will be critical for coping with large
numbers of detectors.

\subsection{Current Status of the Field}
Current experiments have as their first goal detection of the
$E$-modes of polarization.  Table~\ref{pol-t1} provides a summary of all
experiments that are observing, or under construction, or in the
proposal stage.

\input{e6_1_staggs_1023_t2.tex}

{\bf Technology Requirements for Future Polarization Experiments.}
\label{pol-sec4}
Both HEMTs and bolometers are state-of-the-art in the sense of being as
sensitive as is allowed by fundamental limitations. Consequently we cannot
expect substantial improvements in the performance of a single detector
alone to improve the limits on polarization.  The only way to achieve the
sensitivities required is to have many (perhaps tens of thousands) such
detectors.  In addition, the ideal polarization experiment also needs:
\begin{itemize}
\item The ability to integrate for many months without encountering systematic
effects.
\item A wide range of observing frequencies to remove astrophysical
foregrounds from, for example, synchrotron, free-free and dust emission
(see Section~\ref{pol-sec8}).
\end{itemize}

Significant developments are likely in both areas, especially on the
bolometer side, as discussed in Section~\ref{pol-sec3}. The main
barrier in both cases is the cost of development. These technologies are,
in general, beyond the scope of the small groups that have traditionally
performed CMB experiments. The new generation of bolometer
arrays are being built in government labs (JPL, NASA Goddard and NIST, for
example) which are traditionally limited in the types of applications they
can support and which cannot in general support ground-based applications.
As discussed below, ground-based experiments represent the best strategy
for testing various techniques prior to designing a satellite.
Consequently a challenge facing the community is how to fund the
development of these expensive arrays.

{\bf Ground vs Space} A CMB experiment that is
specifically designed to measure $B$-mode
polarization will require significant advances in technology over and
above that which will fly on the Planck Surveyor. Additionally, the
observing regime is completely new in terms of the required sensitivity
and freedom from systematics -- we simply do not yet know how to optimize
an experiment that must be orders of magnitude better than the best
available right now. For this reason, it is prudent to start first with
ground-based experiments which are naturally able to observe for many
months, which are cheap and easy to build, and which can be easily
reconfigured to take advantage of new advances in technology.
Of course, in the near term we also look forward to exciting
results from the balloon-borne B2K (Boomerang with PSBs) and MAXIPOL.

Advocating an initial round of ground-based measurements might seem to
conflict with the traditional ``space is best'' view
in the CMB community.  The time-varying atmospheric emission,
which additionally has spatial structure, hampers ground-based
measurements of the temperature anisotropy.  Spatial switching at
frequencies much higher than the atmospheric time scales has been
used with great success~\cite{mil01}, as has
interferometry~\cite{lei01,padin01}, but the lowest multipoles, where
$B$ modes may dominate, are lost.
 However, the atmospheric emission is
believed to be unpolarized~\cite{keat98}.  Thus, ideal polarimeters
are insensitive to fluctuations in the atmosphere.  In practice,
all polarimeters have some finite response to the common mode (the
intensity $I$), so rejecting atmospheric fluctuations with some form
of spatial switching is likely to be necessary to get residual pickup
down to the sub-$\mu$K level.  Nonetheless, the relative
insensitivity of polarization experiments to atmospheric problems is
a decided advantage.

\subsection{Minimizing Systematic Effects}
\label{pol-sec7}
Measuring even $E$-modes of polarization will require an instrument with
high sensitivity and good control of systematics.  Systematic effects
include the known ones associated with CMB temperature anisotropy
measurements and some new effects that are specific to polarimetry.  A
review of these effects and their effects on bolometric detector systems
can be found in \cite{phil01}.  We summarize the main effects below
(excluding atmospheric emission, discussed above).

{\bf Ground Spillover}. Warm emission from the ground can be reflected
into an instrument by diffraction around the mirrors or scattering from
the mirror support structures, particularly for on-axis systems.
Diffraction is a polarization-dependent process, and thus partially
polarizes the incoming radiation. The most
sensitive measurements of CMB anisotropies to date have either used
interferometric techniques which reduce spillover by shifting it to a
frequency well away from the fringe frequency, or off-axis mirrors which
minimize the blockage of the primary aperture and consequently have very
low spillover. The detectors are fed with carefully designed feed optics
(usually corrugated feeds which have very low sidelobes) that maximize the
illumination of the primary mirror while maintaining very low ground
spillover.

However, off-axis mirrors both generate polarized emission (which will
vary slowly in time unless all mirrors are temperature-controlled),
and increase the degree of instrumental polarization (see below).
For that reason, some groups are returning to the use of
on-axis Cassegrain telescopes,
even though these systems have increased spillover due to the blockage of
the mirrors and a smaller field of view.  Deciding on paper which approach
is best is a difficult exercise. The definitive answer will come from the
next generation of experiments since, as shown in Table~\ref{pol-t1},
both approaches are being adopted by different groups.

{\bf Instrumental (Systematic) Polarization}. Instrumental polarization
occurs when an unpolarized signal at the input of the telescope generates
a polarized signal at the output. In other words, it is a means of
reducing the common mode rejection of the system.
It is usually the result of a mismatch
in the transmission of the two orthogonal polarization modes that are
being differenced, and is generally enhanced by off-axis reflections.

{\bf Cross-polarization}  Cross-polarization occurs if there is cross-talk
between the two orthogonal polarization modes.  The effect is to reduce
the amplitude of a polarized signal and so this effect is also called
polarization efficiency. A well-designed experiment should achieve 1\% or
less cross-polarization.

\subsection{Foregrounds}
\label{pol-sec8}
Foregrounds are potentially a major source of uncertainty in polarization
measurements, just as they are in temperature anisotropy measurements.
Comprehensive reviews of foregrounds and their likely contribution to both
temperature and polarization anisotropies can be found in~\cite{bouch96},
\cite{teg99}. The simulations in~\cite{teg99} show a minimum in the
amplitude of polarized foregrounds at 100GHz, which is why most of the
polarization experiments include this channel. There remains considerable
uncertainty in the amplitude of foregrounds, mainly due to a lack of
polarized maps of large areas of sky at a wide range of frequencies.
Because of this, the ideal experiment will observe at a range of
frequencies to allow foreground removal based on spectral information.

\subsection{Conclusions}
The next few years will see data from a large number of CMB experiments
which will use a variety of different techniques and which will be subject
to different systematic effects. These experiments should measure the
$TE$ cross-correlation and obtain the first estimates of the $E$ mode
power spectrum.  Moreover, there is still a lot that can be accomplished
from the ground and from balloon, and the technology development that will
eventually lead to the ultimate experiment requires us as a community to
move away from the ``traditional'' single university-based experiment.

The ultimate goal of polarization experiments is a measurement of both $E$
and $B$ mode polarization to the limit set by cosmic variance.  This will
require a new generation of experiments with many thousands of background
limited detectors, eventually on a space-based platform.  However, we
conclude that competing design philosophies are best tested on the next
generation of ground and balloon-based experiments.  These experiments are
beginning to approach cost and complexity levels that are too great to
allow the CMB community to continue easily in the few-investigator mode
that has worked so well for the temperature anisotropy measurement
program.  The challenge for our community is to develop collaborations and
sources of funding that will allow us to proceed to the next generation of
CMB measurements.

%% file: e6_1_staggs_1023_t2.tex
\begin{table}[hbtp]
\center
\begin{tabular}{l||l|l|l|l|l}
      &           & Freq./GHz        &       &           & Primary          \\
      & Detectors & (No of focal plane elements) &  Beam & Platform & Mirror
      \\\hline\hline
{\em{\bf Completed:}} & & & & & \\
POLAR      & HEMTs      & 30(1) & 7$^\circ$ & Ground  & None     \\\hline
{\em{\bf Observing:}} & &       &           &         &          \\
CBI Interferometer& HEMTs & 26-36 (13) & 3$'$   & Ground  & On-axis \\
COMPASS    & HEMTs   & 30(1), 90(1)& 20$'$, 7$'$ & Ground & On-axis \\
DASI Interferometer & HEMTs    & 36-46 (13) & 20$'$  & Ground  & None  \\
MAP        & HEMTs    & 22(2), 30(2), 40(4), 60(4), 90(8) & 13$'$--1$^\circ$& Space & Off-axis   \\
PIQUE      & HEMTs    & 40(1), 90(1)& 30$'$, 15$'$ & Ground   & Off-axis
\\\hline
{\em{\bf Under Construction:}} & &       &           &         &          \\
CAPMAP     & HEMTs & 40(4), 90(10) & 6$'$, $3'$ & Ground & Off-axis \\
B2K        & Bolometers & 150(4), 240(4), 350(4) & 12$'$ & Balloon & Off-axis \\
Maxipol    & Bolometers & 150(12), 420(4) & 10$'$ &        Balloon & Off-axis \\
Planck LFI & HEMTs      & 30(2), 44(3), 70(6), 100(17) & 10$'$--33$'$ & Space & Off-axis \\
Planck HFI & Bolometers & 150(4), 220(4), 350(4) & 6$'$ & Space & Off-axis \\
Polatron   & Bolometers & 100(1) & 2.5$'$ & Ground   & On-axis \\
QUEST      & Bolometers & 100(12), 150(24), 220(19) & 6.5$'$-3$'$ & Ground & On-axis \\
AMIBA Interferometer & HEMTs & 90(19) & 2$'$ & Ground & On-axis \\
SPORT      & HEMTs      & 22, 32, 60, 90 & 7$^\circ$ & Space & None
\\\hline
{\em{\bf Proposed:}} & &       &           &         &          \\
Bar-SPORT  & HEMTs & 32, 90 & 30$'$, 12$'$ & Balloon & Unknown \\

POLARBEAR  & Bolometers & 150(3000?)   & 10$'$ & Ground & Off-axis \\
\end{tabular}

\caption[]{Status as of July 2001 of current and planned polarization
experiments.  Note that for the interferometers, each element receives
either right or left circular polarization at any instant.
Also note that for all detectors on Maxipol, and for the detectors at
the two higher frequencies for B2K, each detector receives only one
of the two orthogonal polarization modes.} \label{pol-t1}
\end{table}

%% file: e6_1_staggs_1023_fscmb.tex
\subsection{The Promise of Fine-scale CMB Measurements.}

Measurements of the CMB temperature anisotropy at arcminute angular scales
($\ell > 2000$) can reveal still more about the Universe's fundamental
nature.  The dominant contribution to anisotropies at these fine scales is
not the primordial (or intrinsic) signal reflecting conditions at the
surface of last scattering, but is instead the Sunyaev Zel'dovich (SZ)
effect~\cite{sz72}.   The SZ effect has a frequency signature which allows
it to be disentangled from primordial CMB anisotropies. Other interesting
effects include the generation of distinctive non-Gaussianities by
gravitational lensing, the contribution from CMB photons scattered off
{\it moving} electrons after the epoch of reionization (the
Ostriker-Vishniac effect), and the contribution from the kinetic SZ effect
(see below.)

{\bf Science from the SZ Effect.}
The SZ effect arises from the Compton scattering of CMB photons from
hot electrons in CMB with ionized gas along a line-of-sight to the
surface of last-scattering.  The hottest gas is
located in the potential wells of
rich galaxy clusters where the gas
temperature ranges from 2--15\,keV.  The SZ
effect is {\em independent of the redshift of
the ionized gas} and thus provides a prime technique for detecting
clusters of galaxies out to the epoch of cluster formation.  (Distant
clusters are typically too faint to be seen with X-ray measurements.)
We refer the reader to the report from the Snowmass 2001 P4.1 working
group~\cite{cjk01} for an overview of the exciting cosmology possible from a
large SZ-selected sample of clusters. If a cluster has a peculiar
velocity with respect to the rest
frame of the CMB, the ``kinetic SZ effect'' may also be measured in
principle, allowing estimation of the dark matter content of the
Universe from maps
of the large scale dynamics of the clusters.

Notice that the CMB acts as a uniform backlight
to {\em all} of the hot gas in the universe, including the cooler,
less-dense gas with temperature 8--800 eV which is predicted from
simulations of large scale structure formation to exist as a filamentary
structure between clusters \cite{cen99}.

\begin{figure}
   \includegraphics[width=4.5in]{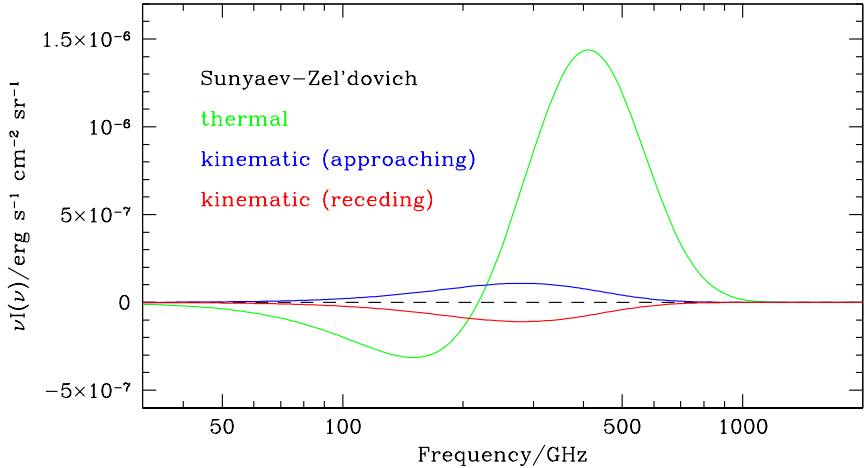}
   \caption[]{Brightness of the SZ effect as a function of frequency. The
solid line is the thermal component, and the two dotted lines show the kinetic
component. The sign of the kinetic component depends on the direction of
the cluster peculiar velocity relative to the observer. The assumed
parameters are $\tau=0.01$, $T_e = 5$~keV, and $v_{\rm pec}$=1000~km/s.}
   \label{sz-fig1}
\end{figure}

{\bf Quantifying the SZ Effect.}
Compton-scattering of CMB photons by the much
hotter electrons in the gas causes a distortion, $\Delta I_{\rm th}$ to
the intensity, $I_{\rm CMB}$, which in the non-relativistic limit is given
by:
\begin{eqnarray}
 \frac{\Delta I_{\rm th}}{I_{\rm CMB}} &=& \frac{x e^x}{(e^x-1)} \left[
x\coth\frac{x}{2} -4\right]y_{\rm th},\mbox{ where} \\
y_{\rm th}&=& \sigma_T \int n_e \frac{kT_{\rm e}}{m_{\rm e} c^2}
{\rm  d}l,
\nonumber
\label{sz-eq1}
\end{eqnarray}
and where $x=h\nu/kT_{\rm CMB}$.  The quantity $y_{\rm th}$ is proportional to
the pressure of the gas integrated along the line-of-sight to the last
scattering surface and depends on $T_{\rm e}$, the temperature of the gas,
$\sigma_T$, the Thompson cross-section, and $n_e$, the electron density.
The distortion characterized by $y_{\rm th}$ is known as the thermal SZ
effect because the amplitude is related to the thermal motions of the
electrons in the clusters.  The thermal SZ effect has a unique spectral
shape (see Fig.~\ref{sz-fig1}) causing a rich cluster to appear as a
``hole'' in the CMB at frequencies $\nu < \nu_{\rm NULL}$, but as a
hot spot at $\nu > \nu_{\rm NULL}$, where the null frequency is
$\nu_{\rm NULL} \sim 217$~GHz.
(Note that
relativistic corrections to Eq.~\ref{sz-eq1} alter the spectrum
slightly so that the null frequency depends weakly on the gas
temperature~\cite{nsi00,reph95}.)

The kinetic SZ effect arises from the bulk motion of the cluster
plasma in the rest frame of the CMB. The change in brightness
is given by
\begin{eqnarray}
\frac{\Delta I_{\rm kin}}{I_{\rm CMB}} &=& \frac{x e^x y_{\rm kin}}{(e^x-1)},
\mbox{ where}\\
 y_{\rm kin} &=&
\sigma_T \int n_e \frac{{\bf v}_{\rm pec}\cdot {\rm d}{\bf l}}{c} = \tau
\frac{v_{\rm pec}}{c}. \nonumber
\label{sz-eq2}
\end{eqnarray}
Here $v_{\rm pec}$ is the mean radial component of the peculiar velocity
of the cluster plasma, ${\bf v}_{\rm pec}$. The optical depth, $\tau$, of
a rich cluster is typically 1\%. The spectral profile of the kinematic SZ
effect is also shown in Figure~\ref{sz-fig1}.  The kinematic effect has
yet to be detected, but for expected cluster peculiar velocities of a few
hundred km/s, it is likely to be at least an order of magnitude fainter
than the thermal effect.

The expressions in Eqs.~\ref{sz-eq1} and~\ref{sz-eq2} give the SZ effect
along a given line of sight through a cluster,  This quantity has {\em no
explicit redshift dependence} because both $\Delta I$ and $I_{\rm CMB}$
scale in the same way with redshift. Consequently, the amplitude and
spectral shape of the SZ effect are independent of the distance to the
cluster. Of course, the redshift does affect the total SZ flux from a
cluster, through geometrical effects~\cite{bbbo96, hc01, vhs01}. These
redshift dependencies are much weaker than for other sources of emission
such as X-ray measurements.

\subsection{Current and Future Experimental Prospects}

Table~\ref{sz-t1} shows current and future experiments intended to
measure the SZ effect.

\input{e6_1_staggs_1023_t3}

{\bf Current Status of SZ Measurements.}
Detections of the thermal SZ effect are now routine (see~\cite{carl98} for a review) and have reached the level where SZ measurements
may  be used as important cosmological probes.

\begin{figure}
   \includegraphics[width=2.5in]{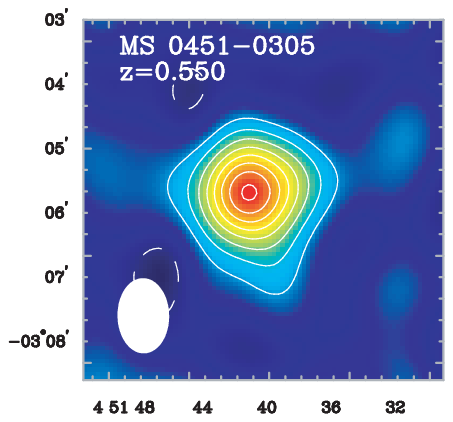} \hspace*{0.5cm}
   \includegraphics[height=2in]{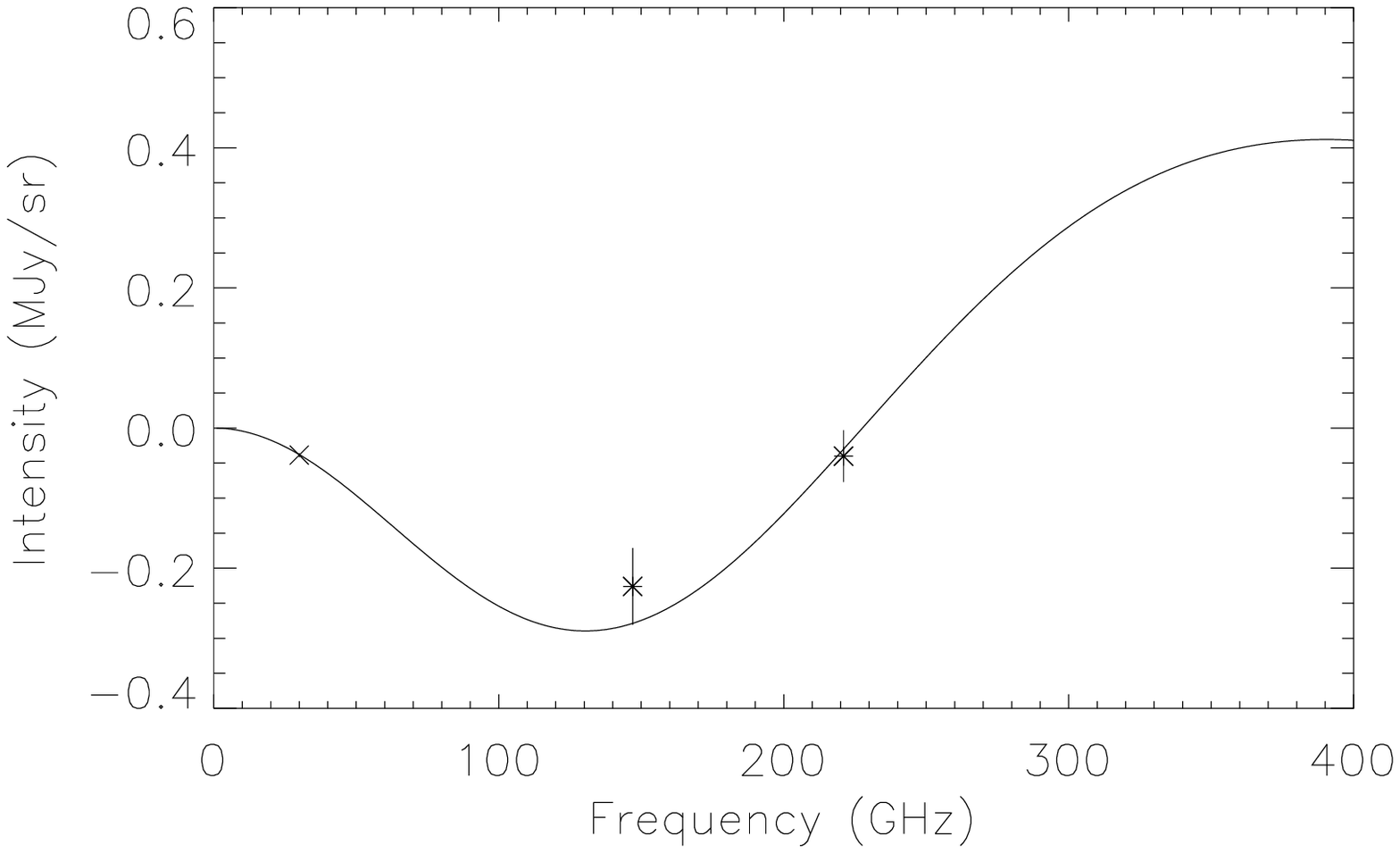}
   \caption[]{Upper Panel: Map of the SZ decrement towards the high redshift cluster MS0451
made using the BIMA array (courtesy J. Carlstrom.).  Lower Panel:
Measurement of the spectrum of MS0451 using the flux measured with the
BIMA array (cross) and the SuZIE experiment (stars).}
   \label{sz-fig4}
\end{figure}

Figure~\ref{sz-fig4} shows high quality data obtained from
mm-wave interferometers outfitted with 30~GHz receivers~\cite{car96}
and the SuZIE instrument. Several low sky coverage surveys to look for clusters have been carried
out.  However to obtain the large numbers of clusters out to a
reasonably small limiting mass (for example, 3-$4\times
10^{14}M_{\cdot}$)  required for setting new constraints on cosmology,
experiments capable of surveying
large portions of the sky to high sensitivity are needed.

{\bf Detection Techniques.} SZ survey instruments fall into two
categories:  HEMT-based interferometers at 30-40~GHz, observing clusters
by their ``SZ decrement,'' and bolometer arrays with multiple frequency
bands straddling the null frequency coupled to large telescopes. Many of
the comparisons between coherent and incoherent techniques made in
Section~\ref{polarization} apply here also.  Atmospheric fluctuations
hamper the bolometer experiments more than the interferometers, but such
fluctuations are smaller at the angular scales of interest than they are
near the peak of the CMB temperature angular power spectrum.  At a good
site (Chile, South Pole, Mauna Kea), the atmosphere will not be a limiting
factor for more than 50\% of the time. An additional consideration here is
that interferometers suffer from not being able to spectrally distinguish
SZ sources; if their resolution is $\ga 2^\prime$, survey mass limits are
set by confusion from the intrinsic CMB anisotropies. On the other hand,
interferometers can make beautiful maps of the clusters (as in
Figure~\ref{sz-fig4}), allowing detailed study of the clusters which will
provide critical guidance for extracting cosmological goals from cluster
surveys.

Therefore, both approaches should be pursued, and, as
Table~\ref{sz-t1} indicates, experiments of both types are already underway.
The table omits comparison of mapping speeds for fear of
misrepresenting the instruments.
Note that mapping speeds are given in units of
deg$^2/(\mbox{K}/\mbox{beam})^2$/s, and
thus quantify how much sky coverage can be mapped to a given
sensitivity per pixel (``beam'') in how much time.
Golwala~\cite{golpri}
notes that the mapping speed for a bolometer array is given by
$\Omega_b N_{det}/{\rm NEy}^2$, while for an interferometer, it is
$\Omega_{\rm FOV}/{\rm NEy}^2$.  Here, $\Omega_b$ and $\Omega_{\rm
FOV}$ are the solid areas of the beam and field of view (FOV), while the noise
equivalent $y$ is ${\rm NEy}=\alpha(\nu){\rm NET}$, with the NET
for a single
detector.  The proportionality $\alpha$ accounts
for the fact that the spectral distortion for the SZ effect from a
cluster is larger at frequencies away from the null.
At 30~GHz, $\alpha\sim 0.0002$, while at 150 GHz, $\alpha\sim 0.35$.
A rule of thumb is that on the ground, bolometer NETs are comparable
to the best HEMT amplifier NETs at 30-40~GHz (to within a factor of two).


{\bf Future Prospects.}
Realizing the full science potential from SZ-selected cluster samples
requires a large number of clusters with redshift information
obtained after their detection in the SZ survey.  Therefore, very
sensitive instruments are required, with large numbers of detectors.
Thus, both future CMB polarization experiments and SZ survey experiments
require development of large multiplexed arrays of bolometers.
Luckily, the technology is coming online already
(e.g. ~\cite{Benford01}), and with continued funding, is likely to be
ready when we are -- after the first data come in from the existing
generation of polarization and SZ-survey instruments.

%% file: e6_1_staggs_1023_t3.tex
\begin{table}[tbh]
\begin{center}
\begin{tabular}{|l|c|c|c|c|c|c|c|r|}
\hline
Name & Type & $N_{det}$ & $\nu$ (GHz) & $N_{ch}$ & FOV & Res & Status & Site/Date\\
\hline
ACBAR~\cite{acbar,holzpri} & Bolo & 16 & 150--345 & 4 &
               $1.5^\circ$ & $5^\prime$ & operating & SP/2001\\
BOLOCAM~\cite{bolocam} & Bolo & 151 & 130--250 & 1 & $\cdots$ &
                $0.6^\prime$--$1.1^\prime$  & operating & CSO/2001\\
AMiBA~\cite{amiba}& Int  & 19 & 90 & 1 & $11^\prime$ &
                $2.6^\prime$ & building & $\cdots$/2003 \\
AMI~\cite{ami} &  Int & 10 & 15 & 1 & $21^\prime$ & $4.5^\prime$ &
                building & UK/2003\\
SZA~\cite{jcpri} & Int & 8 & 30,90 & 2 & $12^\prime,4^\prime$&
               $1.5^\prime$ & building & OVRO/2003\\
ACT/MBAC~\cite{pagepri} & Bolo & 3072     & 145--265 & 3 & $22^\prime$ &
                $0.9^\prime$--$1.7^\prime$  & proposed  & Chile/2004\\
SPT/BA~\cite{jcpri} & Bolo & 1024 & 150-220 & 2 & $1^\circ$
                & $1.3^\prime$ &proposed & SP/2005\\
Planck/HFI~\cite{planck} & Bolo & 50 & 100-580 & 6 &$\cdots$ &
$5^\prime$ & building & space/2007\\
            \hline
\end{tabular}
\caption[]{Comparison of current and planned experiments making SZ cluster
surveys. The second column distinguishes interferometer experiments from
experiments coupling bolometer arrays to large telescopes.  The fifth
column gives $N_{ch}$, the number of frequency channels, while the sixth
and seventh columns give the FWHM of the field of view and the beam
(resolution), respectively.  Note that the ACBAR FOV is obtained by
chopping their bolometer array $3^\circ$ on the sky.   See the text for an
indication of  mapping speeds.   At present, only ACBAR, ACT/MBAC, SPT/BA,
and Planck/HFI plan to collect data at all frequency channels
simultaneously. Not included here is the extended SZA, which adds in the 6
(larger) BIMA telescopes to form a heterogeneous interferometer, providing
for even better removal of radio point sources. }
\label{sz-t1}
\end{center}
\end{table}

%% file: e6_1_staggs_1023_conc.tex
Recent measurements of the CMB temperature anisotropy hang together
with other cosmological data to give a consistent picture of our
universe as spatially flat, dominated by a $\Lambda$-like term, and
with 5-10 times more dark matter than baryons (by mass).

These published CMB data are just the beginning.  In the near future, the
MAP satellite data set will be a gold mine:  full-sky maps in five
frequency bands, with the very stringent control of systematic effects
that only the space environment permits. The Planck data toward the end of
the decade will be even richer, with twice the angular resolution and
broader frequency coverage.

Beyond measuring the primary temperature anisotropy of the CMB lie the next
two exciting frontiers:  the CMB polarization and the fine-scale
structure of the CMB's intensity (including SZ cluster surveys).
Definitive measurements of both
these aspects of the CMB require
experimental sensitivities at least an order of magnitude better than
have yet been achieved.  Individual detectors are near enough
fundamental limits to their sensitivities that the only feasible route
to progress on these two fronts is through large arrays of detectors.
Here, ``large arrays'' is a term encompassing bolometer arrays,
interferometric arrays, and HEMT-receiver-on-a-chip arrays.
Experiments with large arrays require significant funding,
substantial technical development, and therefore,
sizeable collaborations.
At the same time, more modest ``stepping-stone'' experiments will be crucial
for exploring the new systematic effects revealed when the statistical
errors are reduced.  Smaller experiments are faster and more flexible,
and thus can
inform the designs of large-array collaborations in a timely way.

%% file: e6_1_staggs_1023_ack.tex
We would like the thank all the attendees who contributed to this
workshop.  In particular we thank Sunil 
Golwala, Peter Timbie, and Lloyd Knox for allowing us to use materials from their
presentations  herein.